\documentclass[10pt,onecolumn,amsmath,amssymb,floatfix, notitlepage]{revtex4-1}
\usepackage{upgreek}
\usepackage[T1]{fontenc}
\usepackage[utf8]{inputenc}
\usepackage[right=0.75in, left=0.75in, top=1in]{geometry}
\usepackage{microtype,bm,bbm,graphicx,booktabs,times}
\usepackage[usenames,dvipsnames]{xcolor}

\usepackage{setspace}
\selectfont{}
\usepackage{hyperref}
\hypersetup{colorlinks,allcolors=Black,linktoc=all}

\usepackage[capitalize]{cleveref}
\crefformat{section}{Sec.~#2#1#3}

\begin{document}
\title{Data-driven acceleration of photonic simulations}
\author{Rahul Trivedi$^{1,2}$}\email{rtrivedi@stanford.edu}
\author{Logan Su$^{1}$}
\author{Jesse Lu$^{2}$}
\author{Martin F.~Schubert$^{2}$}
\author{Jelena Vuckovic$^{1}$}
\affiliation{$^1$E. L. Ginzton Laboratory, Stanford University, Stanford, CA 94305, USA \\
                $^2$X, Mountain View, CA 94043, USA}
\date{\today}
\begin{abstract}
 Designing modern photonic devices often involves traversing a large parameter space via an optimization procedure, gradient based or otherwise, and typically results in the designer performing electromagnetic simulations of a large number of correlated devices. {In this paper, we investigate the possibility of accelerating electromagnetic simulations using the data collected from such correlated simulations.} In particular, we present an approach to accelerate the Generalized Minimal Residual (GMRES) algorithm for the solution of frequency-domain Maxwell's equations using two machine learning models (principal component analysis and a convolutional neural network). {These data-driven models are trained to predict a subspace within which the solution of the frequency-domain Maxwell's equations approximately lies. This subspace is then used for augmenting the Krylov subspace generated during the GMRES iterations, thus effectively reducing the size of the Krylov subspace and hence the number of iterations needed for solving Maxwell's equations.} By training the proposed models on a dataset of wavelength-splitting gratings, we show an order of magnitude reduction ($\sim 10 - 50$) in the number of GMRES iterations required for solving frequency-domain Maxwell's equations.
\end{abstract}
\maketitle

%%%%%%%%%%%%%%%%%%%%%%%%%%%%%%%%%%%%%%%%%%%%%%%%%%%%%%%%%%%%%%%%%%%%%%%%%%
%%%%%%%%%%%%%%%%%%%%%%%%%%%%%% Section I: Intro %%%%%%%%%%%%%%%%%%%%%%%%%%%%%%%%%%%
%%%%%%%%%%%%%%%%%%%%%%%%%%%%%%%%%%%%%%%%%%%%%%%%%%%%%%%%%%%%%%%%%%%%%%%%%%
\section{Introduction}
Numerical algorithms for solving Maxwell's equations are often required in a large number of design problems, ranging from integrated photonic devices to RF antennas and filters. Recent advances in large-scale computational capabilities have opened up the door for algorithmic design of electromagnetic devices for a number of applications \cite{piggott2015inverse, piggott2014inverse, su2018fully}. {These design algorithms typically explore very high dimensional parameter spaces using gradient-based optimization methods. A single device optimization requires $\sim 500-1000$ electromagnetic simulations, making them the primary computational bottleneck.} However, during such a design process, the electromagnetic simulations being performed are on correlated permittivity distributions (e.g. permittivity distributions generated at different steps of a gradient-based design algorithm). The availability of such data opens up the possibility of using data-driven approaches for accelerating electromagnetic simulations.

Accelerating simulations in a gradient-based optimization algorithm has been previously investigated in structural engineering designs via Krylov subspace recycling \cite{parks2006recycling, wang2007large} --- the key idea in such approaches is to {use the subspace spanned by simulations in an optimization trajectory to augment future simulations being performed in the same trajectory}. {However, these approaches become computationally infeasible if an attempt is made to exploit the full extent of the available simulation data (e.g. use simulations performed in a large number of correlated optimization trajectories) since the augmenting subspace becomes increasingly high-dimensional. Approaches that intelligently select a suitable low-dimensional subspace using machine learning models, like the one explored in this paper, are thus expected to be more efficient at exploiting the available simulation data in practical design settings.} Data-driven methods for solving partial-differential equations have only recently been investigated, with demonstration of learning an `optimal' finite-difference stencil \cite{bar2018data} for time-domain simulations, or using neural networks for solving partial differential equations \cite{han2018solving}.

In this paper, we investigate the possibility of accelerating finite difference frequency domain (FDFD) simulation of Maxwell's equations using data-driven models. Performing an FDFD simulation is equivalent to solving a large sparse system of linear equations, which is typically done using an iterative solver such as Generalized Minimal Residual (GMRES) algorithm \cite{golub2012matrix}. Here we develop an accelerated solver (data-driven GMRES) by interfacing a machine learning model with GMRES. {The machine learning model is trained to predict a subspace that approximates the simulation result, and the subspace is used to augment the GMRES iterations.} Since the simulation is still being performed with an iterative solver, it is guaranteed that the result of the simulation will be accurate --- the performance of the machine learning model only affects how fast the solution is obtained. This is a major advantage of this approach over other data-driven attempts for solving Maxwell's equations \cite{peurifoy2018nanophotonic, liu2018training} that have been presented so far, in which case a misprediction by the model would result in an inaccurate simulation. Using wavelength-splitting gratings as an example, we show an order of magnitude reduction in the number of GMRES iterations required for solving frequency-domain Maxwell's equations.

{This paper is organized as follows - Section \ref{sec:algorithm} outlines the data-driven GMRES algorithm and Section \ref{sec:results} presents results of applying the data-driven GMRES algorithm using two machine learning models, principal component analysis and convolutional neural networks, to simulate wavelength splitting gratings. We show that data-driven GMRES not only achieves an order of magnitude speedup against GMRES, but also outperforms a number of commonly used data-free preconditioning techniques.}

%%%%%%%%%%%%%%%%%%%%%%%%%%%%%%%%%%%%%%%%%%%%%%%%%%%%%%%%%%%%%%%%%%%%%%%%%%
%%%%%%%%%%%%%%%%%%%%%%%%%%%%%% Section II: Model %%%%%%%%%%%%%%%%%%%%%%%%%%%%%%%%%%%
%%%%%%%%%%%%%%%%%%%%%%%%%%%%%%%%%%%%%%%%%%%%%%%%%%%%%%%%%%%%%%%%%%%%%%%%%%
\section{Data-driven GMRES}\label{sec:algorithm}

In the frequency domain, Maxwell's equations can be reduced to a partial differential equation relating the electric field $\textbf{E}(\textbf{x})$ to its source $\textbf{J}(\textbf{x})$:
\begin{align}\label{eq:maxwell_eq}
    \bigg[\nabla \times \nabla \times - \frac{\omega^2}{c^2}\varepsilon(\textbf{x}) \bigg]\textbf{E}(\textbf{x}) = -\textrm{i}\omega \mu_0 \textbf{J}(\textbf{x})
\end{align}
where $\omega$ is the frequency of the simulation, and $\varepsilon(\textbf{x})$ is the permittivity distribution as a function of space. {The finite difference frequency domain method \cite{shin20133d} is a popular approach for numerically solving this partial differential equation -- it discretizes this equation on the Yee grid with perfectly matched layers together and periodic boundary conditions used for terminating the simulation domain to obtain a system of linear equations,} $Af = b$, where $A$ is a sparse matrix describing the operator $\nabla \times \nabla \times - \omega^2 \varepsilon(\textbf{x}) / c^2$ and $b$ is a vector describing the source term $-\textrm{i}\omega \mu_0 \textbf{J}(\textbf{x})$ on the Yee grid.

For large-scale problems, this system of equation is typically solved via a Krylov subspace-based iterative method \cite{golub2012matrix}. {Krylov subspace methods have an advantage that they only access the matrix $A$ via matrix-vector products, which can be performed very efficiently since $A$ is a sparse matrix.
The iterative algorithm that we focus on in this paper is GMRES \cite{golub2012matrix}.} In $i^\text{th}$ iteration of standard GMRES, the solution to $A f = b$ is approximated by $f_i$, where:
\begin{align}\label{eq:gmres}
f_i = 
\underset{f \in  \mathcal{K}_i (A, b)}{\text{argmin}} ||Af - b||^2 
\end{align}
where $\mathcal{K}_i(A, b) = \text{span}(b, Ab, A^2b \dots A^{i-1}b )$ is the Krylov subspace of dimension $i$ generated by $(A, b)$. {The Krylov subspace can be generated iteratively i.e.~if an orthonormal basis for $\mathcal{K}_i(A, b)$ has been computed, then an orthonormal basis for $\mathcal{K}_{i+1}(A, b)$ can be efficiently computed with one additional matrix-vector product. In practical simulation settings, the GMRES iteration is performed till the Krylov subspace is large enough for the residual $r_i = ||Af_i - b|| / ||b||$ to be smaller than a user-defined threshold. We also note that GMRES is a completely data-free algorithm --- it only requires knowledge of the source vector $b$ and the ability to multiply the matrix $A$ with an arbitrary vector.

The number of GMRES iterations can be significantly reduced if an estimate of the solution $A^{-1}b$ is known. To this end, we train a data-driven model on simulations of correlated structures to predict a (low-dimensional) subspace $\mathcal{V}$ within which $A^{-1}b$ is expected to lie. More specifically, the model predicts $N$ vectors, $v_1, v_2 \dots v_N$, such that $\mathcal{V} = \text{span}(v_1, v_2 \dots v_N)$. These vectors can then be used to augment the GMRES iterations (we refer to the augmented version of GMRES as data-driven GMRES throughout the paper) ---} the $i^\text{th}$ iteration of data-driven GMRES can be formulated as:
\begin{align}\label{eq:data_driven_gmres}
f_i = 
\underset{f  \in  \mathcal{V} \oplus \mathcal{K}_i (\tilde{A}, \tilde{b})}{\text{argmin}} ||Af - b||^2
\end{align}
where $\tilde{A}$, and $\tilde{b}$ are given by:
\begin{align}\label{eq:proj_problem}
    \tilde{A} &= P_\perp(Av_1, Av_2 \dots Av_N) A \\
    \tilde{b} &= P_\perp(Av_1, Av_2 \dots Av_N) b
\end{align}
where $P_\perp(Av_1, Av_2 \dots Av_N)$ is the operator projecting a vector out of the space spanned by $Av_1, Av_2 \dots Av_N$. {Note that while in GMRES (Eq.~\ref{eq:gmres}) the generated Krylov subspace is responsible for estimating the entire solution $A^{-1}b$, in data-driven GMRES (Eq.~\ref{eq:data_driven_gmres}) the generated Krylov subspace is only responsible for estimating the projection of $A^{-1}b$ perpendicular to the subspace $\mathcal{V}$. Therefore, a large speedup in the solution of $Af = b$ can be expected if $\mathcal{V}$ is a good estimate of a subspace within which $A^{-1}b$ lies. Moreover, as is shown in the supplement, an efficient update algorithm for data-driven GMRES can be formulated in a manner identical to that formulated for Generalized Conjugate Residual with
inner Orthogonalization and outer Truncation (GCROT) \cite{wang2007large, hicken2010simplified}.} 

\section{Results}\label{sec:results}
\begin{figure}[t]
    \centering
    \includegraphics[scale=0.325]{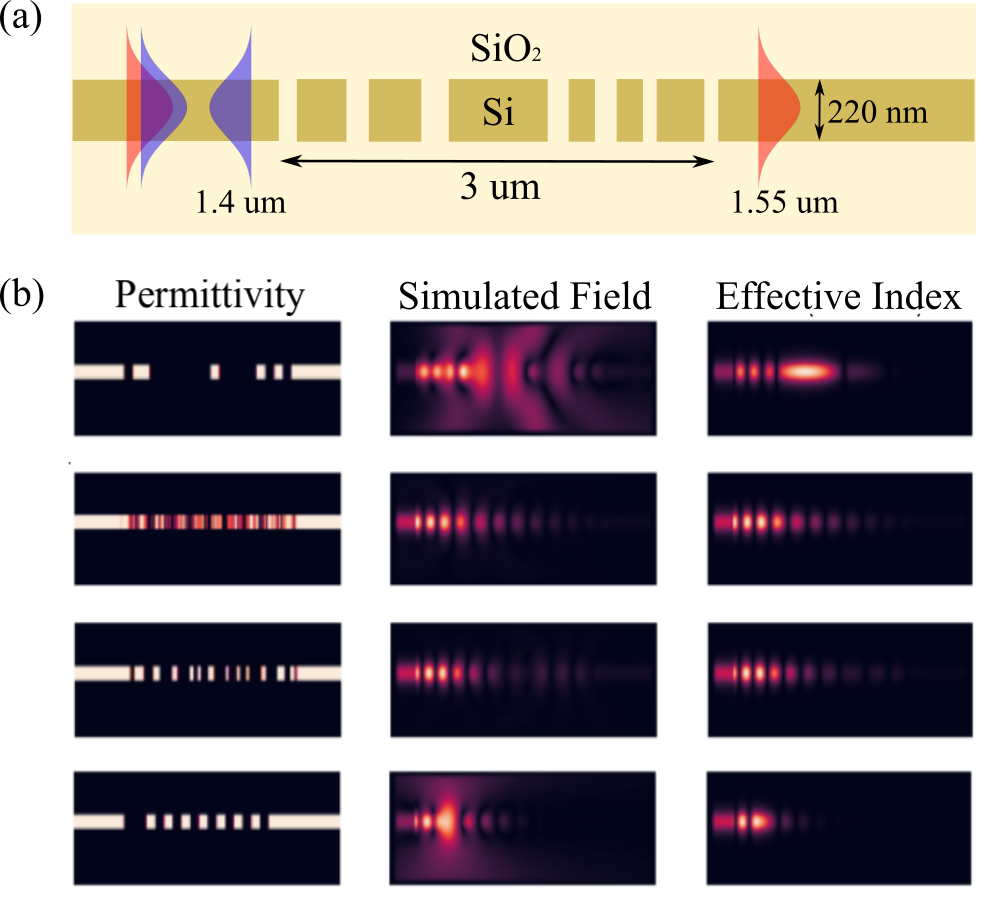}
    \caption{(a) Schematic of the grating splitter device that comprises the dataset.  All the gratings in the dataset are $3~\mu$m long and are designed for a 220 nm silicon-on-insulator (SOI) platform with oxide cladding. {\color{black}We use a uniform spatial discretization of $20$ nm while representing Eq.~\ref{eq:maxwell_eq} as a system of linear equations. The resulting system of linear equations has $229 \times 90 = 20,610$ unknown complex numbers.} (b) Visualizing samples from the dataset --- shown are permittivity distribution, simulated electric fields and effective index fields for 4 randomly chosen samples. All fields are shown at a wavelength of 1.4 $\mu$m. }
    \label{fig:dataset}
\end{figure}

{We investigate two data-driven models to predict the vectors $v_1, v_2 \dots v_N$: principal component analysis and a convolutional neural network. The dataset that we use for training and evaluating these models comprises of a collection of 2D grating splitters [Fig.~\ref{fig:dataset}(a)] which reflect an incident waveguide mode at $\lambda = 1.4 \ \mu$m and transmit an incident waveguide mode at $\lambda = 1.55 \ \mu$m. Throughout this paper, we focus on accelerating simulations of the grating splitters at $\lambda = 1.4\ \mu$m --- so as to train our data-driven models, we provide the dataset with the full simulations of the electric fields in the grating splitter at $1.4\ \mu$m [Fig.~\ref{fig:dataset}(b)]. Additionally, we inuitively expect a well designed data-driven model to perform better if supplied with an approximation to the simulated field as input for predicting the subspace $\mathcal{V}$. To this end, we provide the dataset with effective index simulations \cite{hammer2009effective} of the electric fields in the grating splitter [Fig.~\ref{fig:dataset}(b)]. The effective index simulations are very cheap to perform since they are equivalent to solving Maxwell's equations in 1D, making them an attractive approximation to the simulated field that the data-driven model can exploit. \\}

\begin{figure*}[t]
    \centering
    \includegraphics[width=\textwidth, scale=0.5]{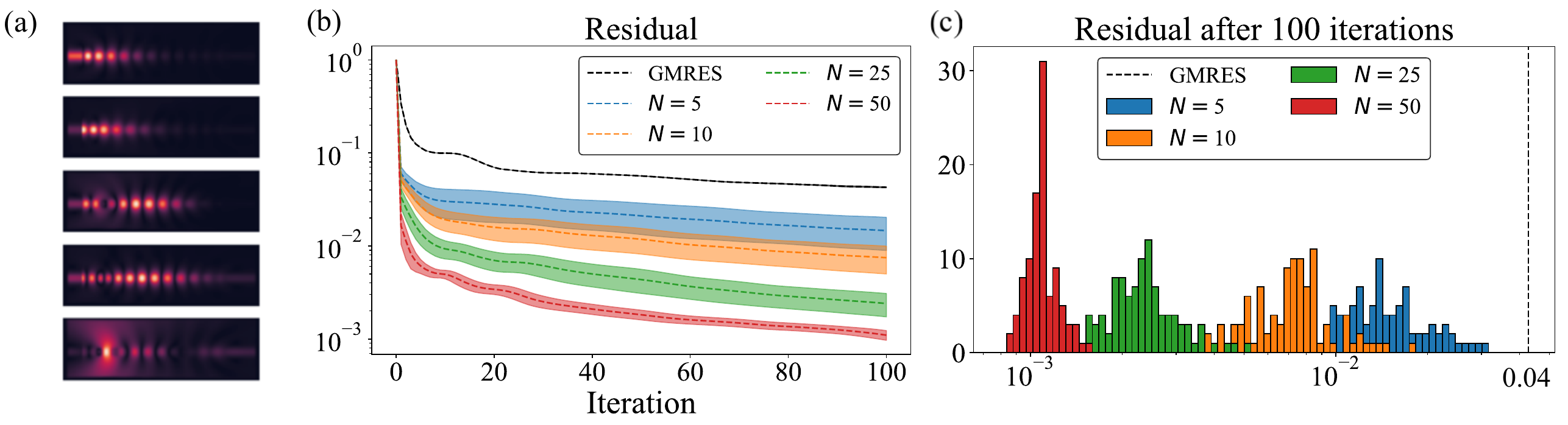}
    \caption{{(a) First five principal components of the electric fields in the grating splitter dataset. (b) Performance of data-driven GMRES on the evaluation dataset when supplied with different number of principal components ($\sim 200$ samples from the training set were used for computing the principal components) --- the dotted line shows the mean residual, and the solid colored background indicates the region within one standard deviation around the mean residual. (c) Histogram of the residual after 100 data-driven GMRES iterations for different $N$ computed over 100 randomly chosen samples from the evaluation dataset. The black vertical dashed line indicates the mean residual after 100 iterations of GMRES over the evaluation dataset.}}
    \label{fig:pca}
\end{figure*}

\noindent\textbf{Principal component analysis:} The first data-driven model that we consider for accelerating FDFD simulations is using the principal components \cite{jolliffe2011principal} computed from the simulated fields in the training dataset as $v_1, v_2 \dots v_N$. The first 5 principal components of the training dataset are shown in Fig.~\ref{fig:pca}(a) --- {we computed these principal components by performing an incomplete singular value decomposition of a matrix formed with the electric field vectors of the training dataset as its columns}. The first two principal components appear like fields ``reflected" from the grating region, whereas the higher order principal components capture fields that are either transmitted or scattered away from the grating devices. Note that the principal components are not necessarily solutions to Maxwell's equations for a grating structure, but provide an estimate of a basis on which the solutions of Maxwell's equations for the grating structures can be accurately represented.

Using principal components as $v_1, v_2 \dots v_N$ in data-driven GMRES, Fig.~\ref{fig:pca}(b) shows the residual $r_i = ||Af_i - b|| / ||b||$ as a function of the number of iterations $i$ and Fig.~\ref{fig:pca}(c) shows the histogram of the residual after 100 iterations over the evaluation dataset. We clearly see an order of magnitude speed up in convergence rate for  GMRES when supplemented with $\geq 5$ principal components. Note that a typical trajectory of data-driven GMRES shows a significant reduction in the residual in the first iteration. This corresponds to GMRES finding the most suitable vector minimizing the residual within the space spanned by $b$ and the supplied principal components. Moreover, the residuals in the data-driven GMRES decrease more rapidly than in GMRES. This acceleration can be attributed to the fact that the Krylov subspace generated corresponds to the matrix $\tilde{A}$ defined in Eq.~\ref{eq:proj_problem} instead of $A$. Since $A$ corresponds to a double derivative operator, its application is equivalent to convolving the electric field on the grid with a $3 \times 3$ filter. Therefore, $A$ generates a Krylov subspace almost on a pixel by pixel basis \cite{ernst2012difficult}. On the other hand, application of $\tilde{A}$ to a vector is equivalent to application of $A$ (which is a $3\times 3$ filter) followed by the projection operator $P_{\perp}(Av_1, Av_2 \dots Av_N)$ (which is a fully dense operator). $\tilde{A}$ can thus generate a Krylov subspace that spans the entire simulation region within the first few iterations, leading to a larger decrease in the residual $r_i$ in data-driven GMRES as compared to GMRES. 

{We also reemphasize the fact that the computation of principal components needs to be performed only once for a given training dataset. Additionally, we observed that we did not need to use the entire dataset (which had $\sim$22,500 electric field vectors) for computing the principal components --- using $\sim200$ randomly chosen data samples already provided a good estimate of the dominant principal components. Consequently, the computational cost of data-driven GMRES is still dominated by the iterative solve which indicates that the residual obtained by data-driven GMRES when compared to GMRES is a good measure of the obtained speedup. \\
}

\noindent\textbf{Convolutional neural network:}
While using principal components achieves an order of magnitude speed up over GMRES, this approach predicts the same subspace $\mathcal{V}$ irrespective of the permittivity distribution being simulated. Intuitively, it might be expected that a data-driven model that specializes $\mathcal{V}$ to the permittivity distribution being simulated would unlock an even greater speed up over GMRES. To this end, we train a convolutional neural network that takes as input the permittivity distribution of the grating device as well as the effective index electric field and predicts the vectors $v_1, v_2 \dots v_N$ [Fig.~\ref{fig:cnn}(a)] that can be used in data-driven GMRES to simulate the permittivity distribution under consideration. {To train the convolutional neural network, we consider two different loss functions:
\begin{enumerate}
\item[(a)] \emph{Projection loss function}: {For the $k^\text{th}$ training example, the projection loss $l^{(k)}_\text{proj}$ is defined by the square of length of the simulated field $f^{(k)}$ that is perpendicular to the space spanned by the vectors $v_1, v_2 \dots v_N$ relative to the square of the length of $f^{(k)}$:
\begin{align}\label{eq:proj_loss}
l^{(k)}_\text{proj} = 
\underset{f  \in  \mathcal{V}}{\text{min}} \ \frac{||f - f^{(k)}||^2}{||f^{(k)}||^2} 
\end{align}
Note that $0 \leq l^{(k)}_\text{proj}\leq 1$, with $l^{(k)}_\text{proj} = 0$ indicating that $f^{(k)}$ lies in the subspace $\mathcal{V}$ and $l^{(k)}_\text{proj} = 1$ indicating that $f^{(k)}$ is orthogonal to $\mathcal{V}$.} 

\item[(b)] \emph{Residual loss function}: The residual loss function can remedy this issue. {For the $k^\text{th}$ training example, the residual loss $l^{(k)}_\text{res}$ is defined by:
\begin{align}\label{eq:res_loss}
l^{(k)}_\text{res} = 
\underset{f  \in  \mathcal{V}}{\text{min}} \ \frac{||A^{(k)}f - b||^2}{||b||^2} 
\end{align}
where $A^{(k)}$ is the sparse matrix corresponding to the operator $\nabla \times \nabla \times - \omega^2 \varepsilon^{(k)} / c^2$ for the $k^\text{th}$ training example. Similar to the projection loss, $0 \leq l_\text{res}^{(k)} \leq 1$, with $l_\text{res}^{(k)} = 0 \implies f^{(k)} = [A^{(k)}]^{-1}b$ lies within the subspace $\mathcal{V}$ and $l_\text{res}^{(k)} = 1\implies f^{(k)} = [A^{(k)}]^{-1}b$ is orthogonal to the subspace $\mathcal{V}$ with respect to the positive definite matrix $W^{(k)} = [A^{(k)}]^\dagger A^{(k)}$ (i.e.~$\langle v, f^{(k)}\rangle_{W^{(k)}} = v^\dagger W^{(k)} f^{(k)} = 0 \ \forall\ v \in \mathcal{V}$).} However, unlike the projection loss function, the residual loss function is an unsupervised loss function i.e.~the simulated electric fields for the structure are not required for its computation.
\end{enumerate}

Since the residual loss function trains the neural network to minimize the residual directly, it accelerates data-driven GMRES significantly in the first few iterations as compared to the projection loss function. This can be clearly seen in Fig.~\ref{fig:cnn}(b), which shows histograms of the residual after 1 iteration and 100 iterations over the evaluation dataset. However, as is seen from the residual vs iteration number plots in Fig.~\ref{fig:cnn}(c), data-driven GMRES with the convolutional neural network trained using the residual loss function slows down after the first few iterations. It is not clear why this slow down happens, and is part of ongoing research. 

\begin{figure*}[b]
    \centering
    \includegraphics[width=\textwidth, scale=0.48]{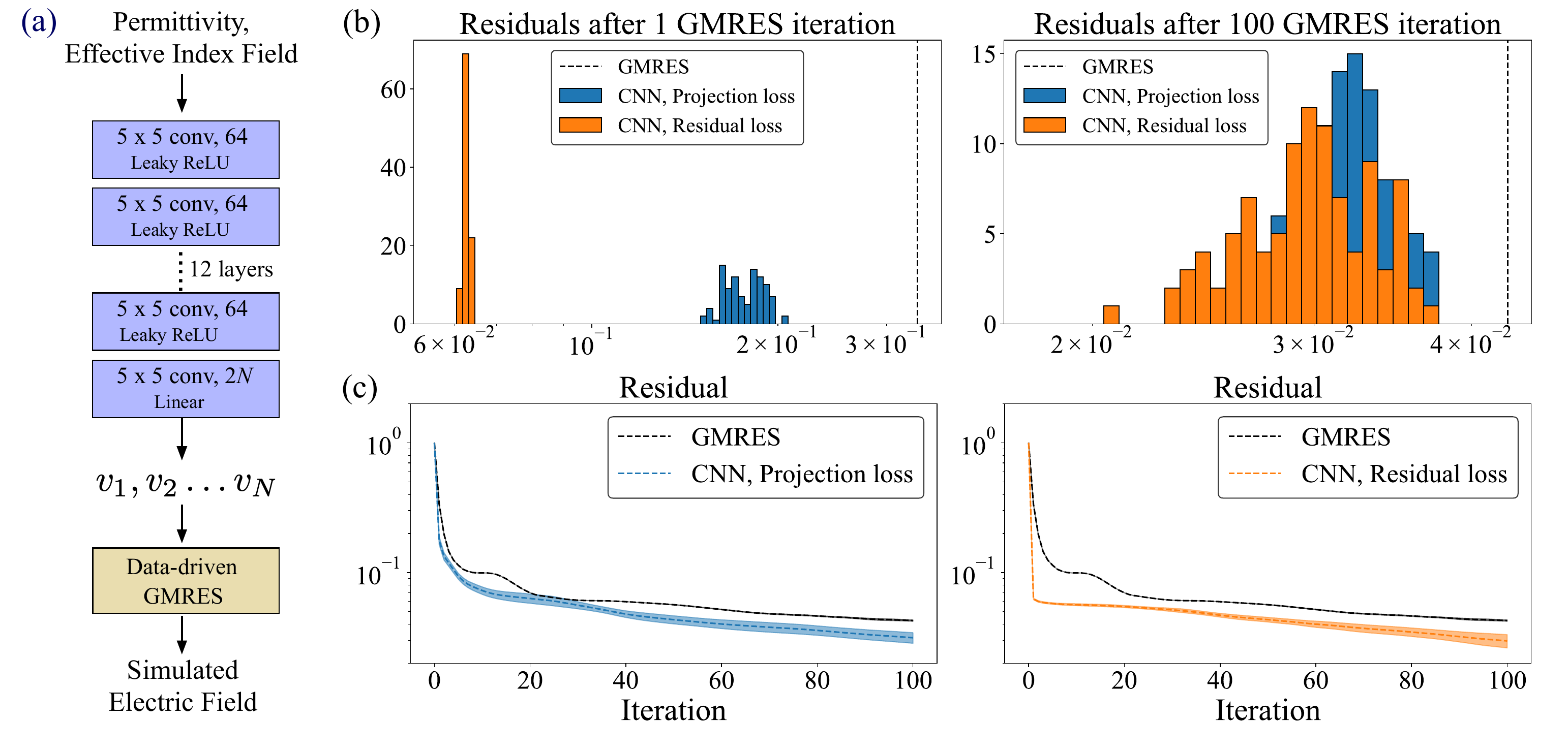}
    \caption{(a) Schematic of the CNN based data-driven GMRES --- a convolutional neural network takes as input the permittivity and effective index field and produces as an output the vectors $v_1, v_2 \dots v_N$. These vectors are then supplied to the data-driven GMRES algorithm, which produces the full simulated field. { (b) Histogram of the residual after 1 and 100 data-driven GMRES iterations evaluated over the evaluation dataset. We consider neural networks trained with both the projection loss function $l_\text{proj}$ and residual loss function $l_\text{res}$. The vertical dashed lines indicate the mean residual after 1 and 100 iterations of GMRES over the evaluation dataset. (c) Performance of the data-driven GMRES on the evaluation dataset when supplied with the vectors at the output of the convolutional neural networks trained with the projection loss function $l_\text{proj}$ and the residual loss function $l_\text{res}$. The dotted line shows the mean residual, and the solid colored background indicates the region within $\pm$standard deviation around the mean residual.}}
    \label{fig:cnn}
\end{figure*}

{We also note that the use of fields obtained with an effective index simulation as an input to the convolutional neural network significantly improves the performance of the convolutional neural network. Empirically, we observed that if the convolutional neural network was trained with the projection loss function but without the effective index fields as an input, the training loss would saturate at $\sim 0.2$, indicating that on an average approximately $20\%$ of the simulated fields $f^{(k)}$ lie outside the subspace $\mathcal{V}$. With a convolutional neural network that takes the effective index fields as an input in addition to the permittivity distribution, the training loss would saturate at a significantly lower value of $\sim 0.05$ indicating that $95\%$ of the simulated field $f^{(k)}$ lies in the subspace $\mathcal{V}$.}}

{ Finally, we point out that the convolutional neural network needs to be trained only once before using it to accelerate GMRES over all the simulations in the evaluation dataset. The results presented in this paper were obtained using a convolutional neural network that required $\sim$10 hours to train, with the training distributed over $\sim$8 GPUs. However, once the neural network was trained, computing the vectors $v_1, v_2 \dots v_N$ for a given structure (i.e. performing the feedforward computation for the neural network shown in Fig.~\ref{fig:cnn}(a)) is extremely fast, and the time take for this computation is negligible when compared to the iterative solve. For full three-dimensional simulations, we expect the iterative solve to be even slower than feedforward computations of most convolutional neural network and consequently believe that data-driven approaches can significantly accelerate electromagnetic simulations in practical settings.} \\

\noindent\textbf{Benchmarks against data-free preconditioning techniques}
{\begin{table}[b!]
  \begin{center}
    \begin{tabular}{ | c | c | c | c | c | c |} % <-- Alignments: 1st column left, 2nd middle and 3rd right, with vertical lines in between
    \hline
      \textbf{Method} & \textbf{Training time } & \textbf{Setup time } & \textbf{GMRES time } & \textbf{Total solve time} & \textbf{Number of iterations}\\
      \hline
      GMRES & $\times$ & $\times$ & 23.34 s & 23.34 s & 115.7 \\
      \hline
      \hline
      PCA: $N = 5$ \  \ & 7.63 s  & $\times$ & 0.87 s & 0.87 s &  6.1 \\
      PCA:   $N = 10$ & 7.63 s & $\times$ & 0.59 s &  0.59 s & 3.5 \\
       PCA:   $N = 25$ & 7.63 s &$\times$ & 0.62 s & 0.62 s & 2.1 \\
       PCA:    $N = 50$ & 7.63 s & $\times$ & 0.98 s & 0.98 s & 2.0 \\
      \hline
       CNN: Projection loss & $\sim$10 hr  & 0.18 s & 13.46 s & 13.52 s & 64.5 \\
       CNN: Residual loss  & $\sim$10 hr & 0.40 s & 13.07 s & 13.47 s & 63.3 \\
       \hline
       \hline
       Jacobi & $\times$ & 0.57 ms & 39.30 s & 39.30 s & 153.8 \\
       \hline
       Gauss Seidel & $\times$ & 17.59 ms & 136.82 s & 136.82 s & 107.3 \\
       \hline
       Ref.~[17] & $\times$ & 5.80 ms & 24.87 s & 24.87 s & 118.7 \\
       \hline
       SOR: $\omega = 0.25$ & $\times$ & 16.40 ms &  170.17 s & 170.17 s & 132.0 \\
       SOR: $\omega = 0.50$ & $\times$ & 17.55 ms & 152.21 s & 152.21 s & 120.2 \\
       SOR: $\omega = 0.75 $ & $\times$ & 17.10 ms & 140.56 s & 140.56 s & 111.7 \\
       SOR: $\omega = 1.25 $ & $\times$ & 18.33 ms & 138.67 s & 138.67 s & 107.5 \\
       SOR: $\omega = 1.50 $ & $\times$ & 20.55 ms & 169.73 s & 169.73 s & 120.4 \\
       SOR: $\omega = 1.75 $ & $\times$ & 21.37 ms & 349.27 s & 349.27s & 235.9 \\
       \hline
       ILU: Drop tol. $=10^{-1}$ & $\times$ & 6.62 s & 13.90 s & 20.52 s & 66.2 \\
       ILU: Drop tol. $=10^{-2}$ & $\times$ & 7.01 s & 4.59 s & 11.60 s & 28.6 \\
       ILU: Drop tol. $=10^{-3}$ & $\times$ &  9.29 s & 0.74 s & 10.03 s & 3.0 \\
       \hline
    \end{tabular}
  \end{center}
  \caption{{Benchmarking data-driven GMRES against data-free preconditioning techniques. The \emph{training time} refers to the amount of time taken to compute the principal components or train the convolutional neural network using the training data. This is done only once, and the same principal components or trained neural network are used for augmenting GMRES over the evaluation data. The \emph{setup time} is the time taken for computing the augmenting vectors or for computing the preconditioner for an evaluation data sample. The {GMRES time} is the time taken for running GMRES to reach a threshold residual of $r_\text{th}=0.04$ over the evaluation data. The \emph{total solve time} is computed by summing the setup time and the GMRES time. The \emph{number of iterations} shown are the number of GMRES iterations required to reduce the residual to below $r_\text{th}=0.04$. The setup time, GMRES time and number of iterations shown in the table are averaged values computed over 50 randomly chosen evaluation data samples. The evaluation times reported are for a machine with 16 CPU cores and 60 GB RAM.}}
    \label{tab:table_benchmarks}
\end{table}}
Accelerating the solution of linear system of equations is a problem that has been studied by the scientific computing community via data-free approaches for several decades. For iterative solvers in particular, using various preconditioners \cite{saad2003iterative} to improve the spectral properties of the system of linear equations has emerged as a common strategy to solve this problem. {In order to gauge the performance of the data-driven approaches outlined in this paper, we benchmark their performance against three stationary preconditioning techniques (Jacobi preconditioner, Gauss-Seidel preconditioner, Symmetric over-relaxation (SOR) preconditioner), incomplete LU preconditioner and a preconditioner designed specifically for finite-difference frequency domain simulations of Maxwell's equations \cite{wonseok} (see the supplement for residual vs.~iteration plots of GMRES with different preconditioners on the evaluation dataset) . Table.~\ref{tab:table_benchmarks} shows the results of these benchmarks:
\begin{enumerate}
\item \emph{Training time}: While using data-driven GMRES augmented with principal components, the training time refers to the amount of time taken to compute the PCA vectors from the training data. While using data-driven GMRES augmented with the output of the CNN, training time refers to the amount of time taken to train the convolutional neural network by minimizing the projection loss function or the residual loss function over the training data. It can be noted that training is done only once, and the same PCA vectors or the same CNN are used for augmenting data-driven GMRES over all the structures in the evaluation dataset.
\item \emph{Setup time}: The setup time refers to the amount of time taken to compute the augmenting vectors or the preconditioner for a given evaluation data sample. Note that there is no setup time while using principal components since they are computed once from the training data. The setup time for the convolutional neural network refers to the amount of time taken to perform the feedforward computation.  The setup time for the preconditioners refers to the time taken for computing the preconditioners.
\item \emph{GMRES time}: This refers to the amount of time taken to run GMRES iterations on the evaluation data sample to reduce the residual below a threshold $r_\text{th}$. The results in Table.~\ref{tab:table_benchmarks} are for $r_\text{th}  = 0.04$ which is chosen to be equal to the mean residual achieved by unpreconditioned GMRES after 100 iterations.
\item \emph{Total solve time}: This is the amount of time taken for performing one simulation. It is calculated as a sum of the setup time (i.e.~time required for calculating the augmenting vectors or the preconditioner) and the GMRES time. We do not include the training time in the solve time since the training is done exactly once for the entire evaluation dataset.
\item \emph{Number of iterations}: This refers to the number of iterations for which GMRES needs to be performed to reduce the residual to below a threshold residual $r_\text{th} = 0.04$.
\end{enumerate}
The numbers presented for setup time, GMRES time and number of iterations are average values obtained by simulating 50 randomly chosen evaluation data samples. We see from Table \ref{tab:table_benchmarks} that data-driven GMRES outperforms every preconditioning technique other than the incomplete LU preconditioner by at least an order of magnitude in terms of total solve time as well as the number of GMRES iterations performed. The best case performance of these preconditioners is only slightly better than unpreconditioned GMRES. Preconditioning partial differential equations describing wave propagation is known to be a difficult problem due to the solution of the wave equations being delocalized over the entire simulation domain even when the source has compact spatial support \cite{ernst2012difficult}. We also note that while the incomplete LU preconditioner does provide a notable speedup over unpreconditioned GMRES, the PCA-based data-driven GMRES outperforms it in terms of the total solve time and CNN-based data-driven GMRES performs comparably to it. Clearly, these benchmarks indicate that the data-driven approaches introduced in this paper have the potential to provide a scalable alternative to data-free preconditioning techniques for accelerating electromagnetic simulations.}

%\begin{figure*}[t]
%   \centering
%    \includegraphics[ scale=0.35]{final_benchmarks.pdf}
%    \caption{{Comparison of the residual obtained after 100 iterations of GMRES (orange), GMRES with different preconditioners (grey), as well as data-driven GMRES with machine learning models (blue) discussed in this paper. The bar graph shows the mean residual over the evaluation dataset, and the error bars indicate the region around the mean corresponding to $\pm$standard deviation in the residual.}}
%    \label{fig:precond_benchmarks}
% \end{figure*}

\section{Conclusion}

In conclusion, we present a framework for accelerating finite difference frequency domain (FDFD) simulations of Maxwell's equations using data-driven models that can exploit simulations of correlated permittivity distributions. We analyze two data-driven models, based on principal component analysis and a convolutional neural network, to accelerate these simulations, and show that these models can unlock an orders of magnitude acceleration over data-free solver. Such data-driven methods would likely be important in scenarios where a large number of simulations with similar permittivity distributions are performed e.g. during a gradient-based optimization of a photonic device.

\section*{Methods}

\noindent\textbf{Dataset:} The grating splitters are designed to reflect an incident waveguide mode at $\lambda = 1.4$ um and transmit an incident waveguide mode at $\lambda = 1.55$ um using a gradient-based design technique similar to that used for grating couplers \cite{su2018fully}. Different devices in the dataset are generated by seeding the optimization with a different initial structure. Note that all the devices generated at different stages of the optimization are part of the dataset. Consequently, the dataset not only has grating splitters that have a discrete permittivity distribution (i.e.~have only two materials -- silicon and silicon oxide), but also grating splitters that have a continuous distribution (i.e. permittivity of the grating splitter can assume any value between that of silicon oxide and silicon). Moreover, the dataset has poorly performing grating splitters (i.e. grating splitters generated at the initial steps of the optimization procedure) as well as well-performing grating splitters (i.e. grating splitters generated towards the end of the optimization procedure). Our dataset has a total of $\sim 30,000$ examples, which we split into a training data set (75$\%$) and an evaluation dataset (25$\%$).\\

\noindent\textbf{Implementation of data-driven models:} Both the data-driven models (PCA and CNN) were implemented using the python library Tensorflow \cite{abadi2016tensorflow}. Any complex inputs (e.g. effective index fields) to the CNN were fed as an image of depth 2 comprising of the real and imaginary parts of the complex input. We use the ADAM optimizer \cite{kingma2014adam} with a batch size of 30 for training the convolutional neural network --- it required $\sim$10,000 steps to train the network.

\bibliography{sample}{}
\end{document}

\title{Supplementary to ``Data-driven acceleration of photonic simulations''}
\author{Rahul Trivedi$^{1,2}$}\email{rtrivedi@stanford.edu}
\author{Logan Su$^{1}$}
\author{Jesse Lu$^{2}$}
\author{Martin F.~Schubert$^{2}$}
\author{Jelena Vuckovic$^{1}$}
\affiliation{$^1$E. L. Ginzton Laboratory, Stanford University, Stanford, CA 94305, USA \\
                $^2$X, Mountain View, CA 94043, USA}
\date{\today}
\maketitle

%%%%%%%%%%%%%%%%%%%%%%%%%%%%%%%%%%%%%%%%%%%%%%%%%%%%%%%%%%%%%%%%%%%%%%%%%%
%%%%%%%%%%%%%%%%%%%%%%%%%%%%%% Section I: Intro %%%%%%%%%%%%%%%%%%%%%%%%%%%%%%%%%%%
%%%%%%%%%%%%%%%%%%%%%%%%%%%%%%%%%%%%%%%%%%%%%%%%%%%%%%%%%%%%%%%%%%%%%%%%%%
\section{Data-driven GMRES}
By following the same steps as in GCROT (Generalized Conjugate Residual with inner Orthogonalization adn outer Truncation), an efficient update rule can be developed for the data-driven GMRES iteration (defined by Eq.~3 of the main text). In this section, we provide more details on the derivation of this update rule.\\

\noindent\emph{Notation and prelimnaries}: 
\begin{enumerate}
    \item The system of equations being solved will be denoted by $Af = b$, with $f$ being the unknown vector being solved for. We also denote by $D$ the size of the system of equations i.e.~$A \in \mathbb{C}^{D\times D}$ and $f, b \in \mathbb{C}^D$. It will be assumed that $A$ is invertible.
    \item Given the vectors $v_1, v_2 \dots v_N$ with which GMRES has to be accelerated (which are assumed to be linearly independent, but not necessarily orthogonal), we will denote by $V$ the matrix that is formed with these vectors as its columns. Note that $V \in \mathbb{C}^{D \times N}$ and $\text{span}(v_1, v_2 \dots v_N) = \text{range}(V)$.
    \item  $\mathcal{K}_n(A, b)$ will denote the Krylov subspace of dimensionality $n$ that is generated by the matrix $A$ and the vector $b$: $\mathcal{K}_n(A, b) = \text{span}(b, Ab, A^2 b \dots A^{n-1}b)$. 
    \item $\tilde{A}$ and $\tilde{b}$ are defined by:
    \begin{subequations}
    \begin{align}
        &\tilde{A} = P_\perp(Av_1, Av_2 \dots Av_N) A \\
        &\tilde{b} = P_\perp(Av_1, Av_2 \dots Av_N) b
    \end{align}
    where $P_\perp(Av_1, Av_2 \dots Av_N)$ is the operator projecting a vector out of $\text{span}(Av_1, Av_2 \dots Av_N)$. For convenience, we will denote this operator by just $P_\perp$. We note that, in general, $\tilde{A}$ is not sparse  even if $A$ is sparse, but for small $N$ multiplication of $\tilde{A}$ with a vector can be computed efficiently by first multiplying the vector by $A$, followed by projecting the resulting vector out of $\text{span}(Av_1, Av_2 \dots Av_N)$.
    \item To conveniently work with $P_\perp$, we perform an incomplete QR decomposition on the matrix $AV$ to obtain an orthogonal matrix $C \in \mathbb{C}^{D \times N}$ and an upper triangular matrix $R \in \mathbb{C}^{N \times N}$: $AV = CR$. It then immediately follows that $P_\perp = \text{I} - CC^\dagger$. Moreover, it is also convenient to precompute and store $R^{-1}$ (Note that if $A$ is invertible, and $v_1, v_2 \dots v_N$ are linearly independent then $R$ is invertible).
    \end{subequations}
\end{enumerate}

\noindent \emph{Arnoldi iteration}: The $i^\text{th}$ iteration of data-driven GMRES approximates the solution to $Af = b$ with $f_i$, where $f_i$ is given by:
 \begin{equation}\label{eq:gmres_iter}
     f_i = 
\underset{f  \in  \text{range}(V) \oplus \mathcal{K}_i (\tilde{A}, \tilde{b})}{\text{argmin}} ||Af - b||^2
 \end{equation}
One of the key ingredients of the GMRES iteration is the Arnoldi iteration which generates an orthonormal basis for the Krylov subspace $\mathcal{K}_{i + 1} (\tilde{A}, \tilde{b})$ from the orthonormal basis for the Krylov subspace $\mathcal{K}_{i}(\tilde{A}, \tilde{b})$. Denoting the orthonormal basis for $K_{i}(\tilde{A}, \tilde{b})$ by $\{q_1, q_2 \dots q_{i}\}$, note that $\text{span}(q_1, q_2 \dots q_i, \tilde{A}q_i) =\mathcal{K}_{i+1}(\tilde{A}, \tilde{b})$. Therefore, $q_{i+1}$ can be computed by orthonormalizing $\tilde{A}q_i$ against $\{q_1, q_2 \dots q_i\}$:
\begin{align}\label{eq:arnoldi}
    q_{i+1} = \frac{v_{i+1}}{||v_{i+1}||}, \ \text{where} \ v_{i+1} = \tilde{A}q_i - \sum_{j=1}^i (q_j^\dagger  \tilde{A}q_i)q_j 
\end{align}
In our implementation, we assume $q_1 = \tilde{b} / ||\tilde{b}||$, and use Eq.~\ref{eq:arnoldi} to generate $q_2, q_3 \dots$ and so on. Note that $q_i \perp \text{span}(Av_1, Av_2 \dots Av_N) \ \forall \ i$, or equivalently $C^\dagger Q_i = 0 \ \forall \ i$. Denoting by $Q_i$ the matrix formed with the vectors $q_1, q_2 \dots q_i$ as its columns ($Q_i \in \mathbb{C}^{D \times i}$), the Arnoldi iteration can be expressed as the following relationship between $Q_{i+1}$ and $Q_i$:
\begin{align}\label{eq:arnoldi_matrix}
    \tilde{A}Q_i = Q_{i+1}H_{i, i + 1} \implies A Q_i = Q_{i + 1} H_{i, i + 1} + CC^\dagger A Q_i
\end{align}
where $H_{i, i+1} \in \mathbb{C}^{(i + 1) \times i}$ is an upper Hessenberg matrix defined by:
\begin{align}
    H_{i, i + 1} =
    \begin{bmatrix}
    q_1^\dagger \tilde{A} q_1 & q_1^\dagger \tilde{A} q_2 & q_1^\dagger \tilde{A} q_3 & \dots & q_1^\dagger \tilde{A}q_i \\
    ||v_2|| & q_2^\dagger \tilde{A} q_2 & q_2^\dagger \tilde{A}q_3 & \dots & q_2^\dagger \tilde{A}q_i \\
    0 & ||v_3|| & q_3^\dagger \tilde{A}q_3 & \dots & q_3^\dagger \tilde{A}q_i \\
    0 & 0 & ||v_4|| & \dots & q_4^\dagger \tilde{A}q_i \\
    \vdots & \vdots & \vdots & \ddots & \vdots \\
    0 & 0 & 0 & \dots & ||v_{i + 1}||
    \end{bmatrix}
\end{align}

\noindent\emph{Calculating $f_i$}: Consider now solving the optimization problem in Eq.~\ref{eq:gmres_iter}. Since the optimization variable $f$ is in the space $\text{range}(V) \oplus \mathcal{K}_{i}(\tilde{A}, \tilde{b})$, it can be expressed as:
\begin{align}
    f = VR^{-1}x + Q_i y
\end{align}
where $x \in \mathbb{C}^{N}$ and $y \in \mathbb{C}^{i}$. Thus, it follows that:
\begin{align}\label{eq:res_simp}
    ||A f - b ||^2 &= || AVR^{-1}x + A Q_i y - b||^2 \nonumber\\
                   &= \bigg| \bigg| [ C \ Q_{i + 1}] \begin{bmatrix} I & C^\dagger A Q_i \\ 0 & H_{i, i + 1} \end{bmatrix} \begin{bmatrix} x \\ y \end{bmatrix} - b \bigg| \bigg|^2
\end{align}
wherein we have used $AV = CR$ and Eq.~\ref{eq:arnoldi_matrix}. Note that since $C$ and $Q_{i + 1}$ are independently orthogonal matrix, and $C^\dagger Q_{i + 1} = 0$, it follows that $[C \ Q_{i + 1}]$ is an orthogonal matrix. Eq.~\ref{eq:res_simp} can now be further simplified to:
\begin{align}
    ||Af - b||^2 = \bigg| \bigg|  \begin{bmatrix} I & C^\dagger A Q_i \\ 0 & H_{i, i + 1} \end{bmatrix} \begin{bmatrix} x \\ y \end{bmatrix} - \begin{bmatrix} C^\dagger b \\ Q_{i + 1}^\dagger b\end{bmatrix} \bigg| \bigg|^2 + ||(\text{I} - CC^\dagger - Q_{i+1}Q_{i + 1}^\dagger) b ||^2
\end{align}
Therefore, $f_i = VR^{-1} x_i + Q_i y_i$, where
\begin{align}\label{eq:final_lsq}
    x_i, y_i = \underset{x, y}{\text{argmin}} \ \bigg| \bigg|  \begin{bmatrix} I & C^\dagger A Q_i \\ 0 & H_{i, i + 1} \end{bmatrix} \begin{bmatrix} x \\ y \end{bmatrix} - \begin{bmatrix} C^\dagger b \\ Q_{i + 1}^\dagger b\end{bmatrix} \bigg| \bigg|^2
\end{align}
We have thus reduced the problem of calculating $f_i$, which was a constrained least squares problem, to an unconstrained least squares problem (Eq.~\ref{eq:final_lsq}) of size $i + N$, which can be solved numerically (e.g.~using QR factorization).

\section{Benchmarks for data-free preconditioners}\label{sec:precond}
\begin{figure}[t]
    \centering
    \includegraphics[scale=0.32]{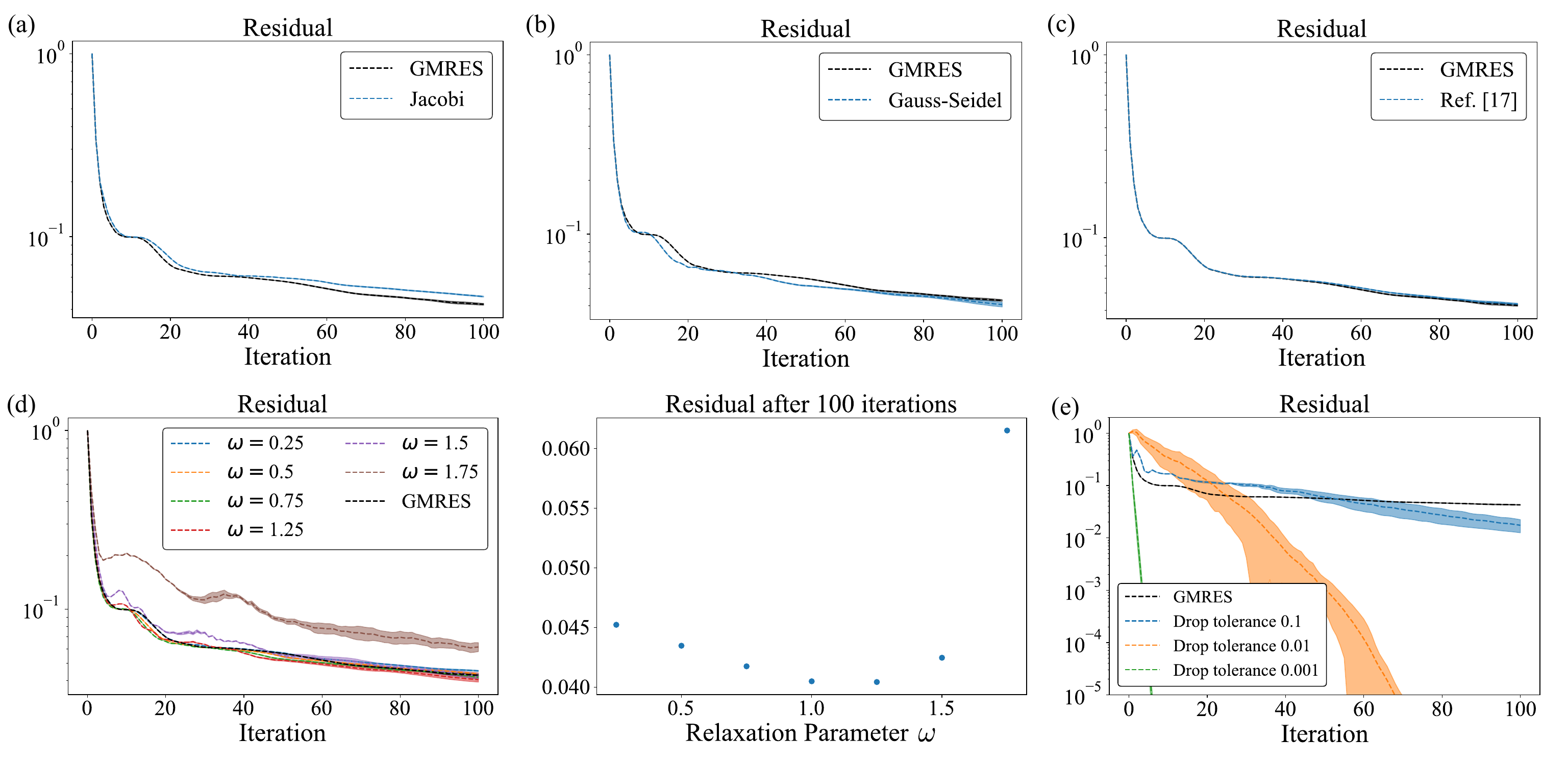}
    \caption{Performance of the different preconditioners described in section ~\ref{sec:precond} on the evaluation dataset. (a) Jacobi preconditioner, (b) Gauss-Siedel preconditioner, (c) preconditioner from ref.~[17] of main text (d) Symmetric over-relaxation (SOR) preconditioner for different relaxation parameter $\omega$ and (e) Incomplete LU preconditioner for different drop tolerances.}
    \label{fig:precond}
\end{figure}
Here we present the results of applying some data-free preconditioners on the simulation problem. Given a left preconditioner $P_L$ and a right preconditioner $P_R$, the system of equations being solved is transformed from $A f = b$ to $A'f' = b'$ where:
\begin{align}
    A' = P_L A P_R, \ b' = P_L b \ \text{and} \ f' = P_R^{-1}f
\end{align}
We study the following four preconditioners:
\begin{enumerate}
    \item \emph{Jacobi preconditioner}: The Jacobi preconditioner \cite{saad2003iterative} is given by:
    \begin{align}
        P_L = \mathcal{D}(A)^{-1} \ \text{and} \ P_R = I
    \end{align}
    where $\mathcal{D}(A)$ is a diagonal matrix formed from the diagonal entries of the matrix $A$. The performance of Jacobi preconditioner on the evaluation dataset is shown in Fig.~\ref{fig:precond}(a).
    \item \emph{Gauss-Siedel preconditioner}: The Gauss-Siedel preconditioner is given by:
    \begin{align}
        P_L = [\mathcal{D}(A) + \mathcal{L}(A)]^{-1} \ \text{and} \ P_R = I
    \end{align}
    where $\mathcal{L}(A)$ is a strictly lower-triangular matrix formed by the elements of $A$ below the main diagonal. Note that application of this preconditioner requires the solution a lower triangular system of equations. The performance of the Gauss-Siedel preconditioner on the evaluation dataset is shown in Fig.~\ref{fig:precond}(b).
    \item \emph{Preconditioner from ref. [17]}: This preconditioner is specific to Maxwell's equations. $P_R$ and $P_L$ are diagonal matrices constructed from the grid spacing (including the complex stretching due to PMLs) in the simulation domain. Details of this preconditioner can be found in ref. [17] for main text. The performance of this preconditioner on the evaluation dataset is shown in Fig.~\ref{fig:precond}(c).
    \item \emph{Symmetric over-relaxation (SOR) preconditioner}: The SOR preconditioner \cite{saad2003iterative} is given by:
    \begin{align}
        P_L = [\mathcal{D}(A) + \omega \mathcal{L}(A)]^{-1} \ \text{and} \ P_R = I
    \end{align}
    where $\mathcal{L}(A)$ is a strictly lower-triangular matrix formed by the elements of $A$ below the main diagonal and $\omega$ is a tunable parameter (referred to as the relaxation parameter) in the preconditioner which can be between 0 to 2. Note that the SOR preconditioner for $\omega = 1$ is identical to the Gauss-Siedel preconditioner. Additionally, application of the SOR preconditioner requires the solution of a lower triangular system of equations. The performance of the SOR preconditioner on the evaluation dataset is shown in Fig.~\ref{fig:precond}(d) --- we see that the best performance of SOR preconditioner on our dataset is achieved for $\omega = 1.25$.
    
\item \emph{Incomplete LU}: This preconditioner seeks an upper and lower triangular matrix, $U$ and $L$ such that the product $LU$ is approximately equal to the matrix $A$ \cite{saad2003iterative}. The preconditioner is then given by:
\begin{align}
P_L = U^{-1}L^{-1} \ \text{and} \ P_R = I
\end{align}
The deviation of the $LU$ from $A$ is controlled with a drop tolerance parameter --- a larger drop tolerance implies a faster computation of $L$ and $U$ but a worse approximation to $A$ and therefore a less useful preconditioner. The performance of incomplete LU preconditioner on the evaluation dataset is shown in Fig.~\ref{fig:precond}(e) for drop tolerances of $0.1, 0.01$ and $0.001$.
\end{enumerate}

\bibliography{sample}{}